# Using Machine Learning to Forecast Market Direction with Efficient Frontier Coefficients

Nolan Alexander, William Scherer

**Nolan Alexander**
is student in Systems Engineering, University of Virginia, Charlottesville, VA
nka5we@virginia.edu

**William Scherer**
is Professor of Systems Engineering, University of Virginia, Charlottesville, VA
wts@virginia.edu

# Using Machine Learning to Forecast Market Direction with Efficient Frontier Coefficients

February 2023

**ABSTRACT**

We propose a novel method to improve estimation of asset returns for portfolio optimization. This approach first performs a monthly directional market forecast using an online decision tree. The decision tree is trained on a novel set of features engineered from portfolio theory: the efficient frontier functional coefficients. Efficient frontiers can be decomposed to their functional form, a square-root second-order polynomial, and the coefficients of this function captures the information of all the constituents that compose the market in the current time period. To make these forecasts actionable, these directional forecasts are integrated to a portfolio optimization framework using expected returns conditional on the market forecast as an estimate for the return vector. This conditional expectation is calculated using the inverse Mills ratio, and the Capital Asset Pricing Model is used to translate the market forecast to individual asset forecasts. This novel method outperforms baseline portfolios, as well as other feature sets including technical indicators and the Fama-French factors. To empirically validate the proposed model, we employ a set of market sector ETFs.

**Highlights**

- We forecast the direction of the market with an online decision tree using a novel feature set and present a framework to incorporate the forecasts to portfolio optimization. We present empirical results with a set of market sector ETFs.
- We present a novel feature set for a machine learning framework based on portfolio theory: the efficient frontier coefficients. Efficient frontiers can be defined by three interpretable coefficients that contain information about the market constituents.
- The directional forecasts can be integrated into a portfolio optimization framework by estimated the return vector using conditional expectations of the market's direction with the inverse Mills ratio and the Capital Asset Pricing Model.

The features that we engineered to forecast the market, the efficient frontier coefficients, have their foundation in Modern Portfolio Theory (MPT). MPT is a framework developed by Markowitz (1952) for optimally allocating portfolio weights through mean-variance optimization. Mean-variance optimization is a quadratic programming problem of allocating portfolios that minimize volatility, while generating a fixed expected return. Solving this optimization problem at multiple expected returns can allow users to construct a pareto frontier consisting of optimal portfolios, known as an efficient frontier. The efficient frontier provides a visualization of the risk-return trade-off for different portfolios.

Since the efficient frontier includes a continuous set of possible weights to choose from, an important problem is selecting the best portfolio on the efficient frontier to invest in. An investor would desire a portfolio that can generate an excess return to the risk-free rate with low volatility. The Sharpe ratio is a ratio of this excess return and standard deviation, and is a common portfolio metric (Sharpe 1964). The Capital Market Line (CML) is a line corresponding with the market portfolio and risk-free rate, with a slope of the market portfolio Sharpe ratio under the Capital Asset Pricing Model (CAPM) framework. The portfolio on the efficient frontier that is tangent to the CML, known as the tangency portfolio, therefore maximizes the Sharpe ratio, and is the theoretical best portfolio on the efficient frontier (Tobin 1958).

The tangency portfolio model makes several assumptions, but we will focus only on the assumption that the market is stationary. The market is clearly nonstationary in practice, as future expected asset returns and covariances will not be identical to the asset's historical data. This assumption frequently leads to significant estimation errors in practice, which limits the performance of the mean-variance model. Several researchers have already developed models that attempt to relax this assumption.

Black and Litterman (1992) proposed a Bayesian model to provide superior posterior estimates of the return vector and covariance matrix. The Black-Litterman model uses two priors: investors' beliefs on the asset movements and the market capitalized equilibrium view, which weights the assets by their market capitalization. To use the model, investors must provide several matrices and vectors as inputs including views on assets. Black and Litterman derived a closed-form solution to the model that calculates the posterior parameter estimates implied by the views.

While the Black-Litterman model can improve estimation of the MPT parameters, there are still limitations. The Black-Litterman model can be difficult to use in practice because it requires a large number of estimates including matrices and vectors. The level-of-unconfidence can be particularly difficult to estimate because investors often have difficulty quantifying the confidence of the views they estimate. Investors also have difficulty interpreting and trusting these models relative to the mean-variance model because of their complexity.

While both parameters are important to estimate, the return vector is the parameter that the mean-variance model is often more sensitive to. In addition, the return vector is particularly difficult to estimate because the autocorrelation of asset returns is often insignificant expect for high-frequency transactions (Cont 2001).

Separate from portfolio optimization, other researchers have developed machine learning models to forecast market returns. A number of researchers trained SVMs, Random

Forest, boosted trees, and Neural Networks models on technical indicators to forecast the intraday, daily, weekly, and monthly return direction of the S&P 500 (Kumar et al. 2006), Patel et al. 2015). In addition to forecast the entire market, other researchers have trained these models on technical and fundamental indicators to forecast individual stock return directions (Krauss et al. 2017, Sirignano and Cont 2019, Aldridge and Avellaneda 2019, Kim et al. 2022). In addition to forecasting the direction of the stocks, other researchers have used machine learning models to forecast stock returns as continuous data (Rapach et al. 2018, Wang et al. 2021, Li et al. 2022).

While these researchers used technical indicators, which can be highly correlated to momentum, and fundamental data, we instead propose using the set of efficient frontier coefficients. In addition, these works focus on using more complex, and less interpretable machine learning models to achieve higher performance, but we will use a simpler model: a classification and regression tree (CART). A CART model is more interpretable because it only generates a single tree, so after a forecast, we can follow that tree to see why it made a certain prediction. Rather than focusing on the prediction metrics such as accuracy, similar to these previous works, we integrate our forecasts to a mean-variance portfolio, and observe the portfolio metrics.

We provide a novel approach to help relax the MPT assumption of stationarity by providing improved estimates of the return vector as expected returns conditional on market forecasts. Each month, we train a CART decision tree on the lagged monthly efficient frontier coefficients using a 5-fold time-series split to tune the hyperparameters, then forecast whether the S&P 500 will be up or down the next month. With this forecast, we translate the impact of the forecast on each asset in the portfolio using CAPM. Then, we use the expected returns conditional on the CART forecast, calculated with the inverse Mills ratio, as our estimate of the return vector. Using this estimate of the return vector, we compute the tangency portfolio. We then walk the model forward one month, and repeat for the remaining out-of-sample time period.

This novel approach is related to three previous works. A common method for creating an efficient frontier is to use mean-variance optimization at several different returns and extrapolate between these, but Merton (1972) derived a closed-form solution of the efficient frontier as a square root second-order polynomial function with three coefficients. Previously, Alexander et al. (2021) developed a model to forecast these coefficients to provide improved parameter estimates. Alexander and Scherer (2023) then defined a set of novel interpretable coefficients that can be used to forecast tangency portfolios to provide improved out-of-sample Sharpe ratios, and developed a robust forecasting and weights extraction model. Rather than forecasting the coefficients, this paper will focus on using the coefficients as predictors to forecast the market.

This paper provides two main contributions. (1) This paper provides a set of feature-engineered predictors that can be used to forecast the market with CART, the set of interpretable efficient frontier coefficients: $r_{MVP}$, $\sigma_{MVP}$, and $u$. (2) This paper provides a method to integrate directional market forecasts to a portfolio framework using CAPM and the conditional expectations of the returns given the market forecast. We provide empirical results demonstrating that this novel approach yields superior Sharpe ratio relative to three benchmarks. This superior performance helps validate the proposed efficient frontier coefficients tree structure of the market.

## ASSETS DATA

To test the proposed models, we used a diverse set of ETF assets corresponding to the S&P market sectors. We included all 9 original S&P Sector ETFs (Technology, Health Care, Finance, Energy, Materials, Consumer Discretionary, Industrial, Utilities, and Consumer Staples). The 24 years of assets data from 1999 to 2022 were collected using the Yahoo Finance API. Also, for the risk-free rate of the tangency portfolios, we used the French-Fama 3-factor data from Dr. French's website (2021).

## FORMAL NOTATION OF MODERN PORTFOLIO THEORY

### Mean-variance Optimization

The following is the formal notation used in MPT: $r_i$ is the expected ln return of an asset in $\boldsymbol{r}$. $\boldsymbol{w}_i$ is the weight of asset $i$, satisfying $\sum_{i=1}^{n} w_i = 1$. $r_{target} = \sum_{i=1}^{n} \boldsymbol{w_i r_i}$ is the target expected return of the portfolio. $\boldsymbol{V}$ is the covariance matrix. $\boldsymbol{e} = (1, \dots, 1)^T$. The formulation of the mean-variance optimization is as follows:

$$\min_{\boldsymbol{w}} \boldsymbol{w}^T \boldsymbol{V} \boldsymbol{w}$$
$$\text{s.t. } \boldsymbol{w}^T \boldsymbol{r} = r_{target} \text{ and } \boldsymbol{e}^T \boldsymbol{w} = 1$$

By adding a constraint that $\boldsymbol{w}_i \geq 0 \ \forall i$, we can prevent shorting, but this paper uses models that do allow shorting to better represent a breadth of investment strategies. We used the Sequential Least-squares Quadratic Programming algorithm (SLSQP) in the SciPy package of Python to perform the mean-variance optimization.

### The Tangency Portfolio

The theoretical best portfolio on the efficient frontier is the one that intersects the Capital Market Line (CML) under the Capital Asset Pricing Model (CAPM), which is known as the tangency portfolio (Sharpe 1966). The tangency portfolio maximizes a common portfolio metric, the Sharpe ratio, which is defined by $\frac{r_p - r_f}{\sigma_p}$. The standard approach to determining the tangency portfolio (Luenberger 1997) frames it as an optimization problem:

$$\max_{\boldsymbol{w}} \frac{\boldsymbol{w}^T \boldsymbol{r} - r_f}{(\boldsymbol{w}^T \boldsymbol{V} \boldsymbol{w})^{1/2}}$$

### Efficient Frontiers Coefficients

Using Lagrange multipliers, Merton (1972) derived a functional representation of the efficient frontier as a square root second-order polynomial with three coefficients (Petters & Dong 2016):

$$A = \boldsymbol{e}^T \boldsymbol{V}^{-1} \boldsymbol{e} > 0 \qquad (1)$$
$$B = \boldsymbol{r}^T \boldsymbol{V}^{-1} \boldsymbol{e}$$
$$C = \boldsymbol{r}^T \boldsymbol{V}^{-1} \boldsymbol{r} > 0$$

The closed-form equation for the efficient frontier is

$$\sigma(r) = \sqrt{\frac{Ar^2 - 2Br + C}{AC - B^2}} \tag{2}$$

## A NOVEL SET OF INTERPRETABLE EFFICIENT FRONTIER COEFFICIENTS

Alexander and Scherer (2023) derived an equation with the same form as (2), but using more interpretable coefficients, as

$$\sigma^2(r) = (u^{-1}(r - r_{MVP}))^2 + \sigma_{MVP}^2 \tag{3}$$

$r_{MVP}$ and $\sigma_{MVP}$ are the return and standard deviation of the minimum variance point (MVP). $u$ is the rate of curvature of the efficient frontier quadratic utility function. An efficient frontier with a greater $u$ diminishes more slowly, and therefore has better trade-offs of risk and return at every efficient portfolio except for the minimum variance point. $u$ can be interpreted as the importance of diversifying in the current market, and we provide justification for this interpretation in the following section. We can calculate each of these more interpretable coefficients as functions of $A, B, C$ by rearrange terms as

$$r_{MVP} = \frac{B}{A}, \sigma_{MVP} = \frac{1}{\sqrt{A}}, u = \sqrt{\frac{AC - B^2}{A}} \tag{4}$$

Each of these interpretable coefficients control one graphical component of the efficient frontier: each coordinate of the vertex, and the rate of curvature. An increase in $r_{MVP}$ indicates greater returns for all levels of risk, which would imply an improving market. An increase in $\sigma_{MVP}$ indicates greater risk for all returns, which would imply a worsening, more volatile market. An increase in $u$ would signal a better market for risk-seeking investors.

### Interpretation of the Utility Coefficient u

While $r_{MVP}$ and $\sigma_{MVP}$ are easy to financially interpret as they directly correspond to MVP risk and return, $u$ is less so. To better understand $u$, Alexander and Scherer (2023) previously derived an alternate formulation of $u$. Starting with Equation (4) and Equation (1), we have

$$u^2 = \frac{AC - B^2}{A} = \frac{(\boldsymbol{e}^T\boldsymbol{V}^{-1}\boldsymbol{e})(\boldsymbol{r}^T\boldsymbol{V}^{-1}\boldsymbol{r}) - (\boldsymbol{r}^T\boldsymbol{V}^{-1}\boldsymbol{e})^2}{\boldsymbol{e}^T\boldsymbol{V}^{-1}\boldsymbol{e}}$$

We can simplify the expression (see Appendix A) into

$$u = \sqrt{\boldsymbol{r}^T\boldsymbol{V}^{-1}\boldsymbol{r}} \cdot \sqrt{1 - S_c(\boldsymbol{r}, \boldsymbol{e})^2}$$

Therefore, $u$ can be defined by the product of the Mahalanobis Distance of $\boldsymbol{r}$ to $\vec{\boldsymbol{0}}$, $\sqrt{\boldsymbol{r}^T\boldsymbol{V}^{-1}\boldsymbol{r}}$, and a function of the cosine similarity in the covariance-weighted inner product space between the return vector and the ones vector, $\sqrt{1 - S_c(\boldsymbol{r}, \boldsymbol{e})^2}$.

The Mahalanobis distance is in effect a generalization of the notion of an L2-normed Z-score to a multidimensional distribution, using the covariance matrix instead of the one-dimensional standard deviation. $\sqrt{\boldsymbol{r}^T\boldsymbol{V}^{-1}\boldsymbol{r}}$ measures how statistically significantly far $\boldsymbol{r}$ is from $\vec{\boldsymbol{0}}$, accounting for covariance. An investor would prefer a greater Mahalanobis distance because the expected returns would be higher relative to the risk measured by the standard deviations and cross-correlations.

The $\sqrt{1-S_c(\boldsymbol{r},\boldsymbol{e})^2}$ is a function of the cosine similarity, $S_c$, which is the cosine similarity in the covariance-weighted inner product space, measuring similarity between two vectors as the cosine of the angle between them under this inner product. This function of cosine similarity is 0 when each of the returns in $\boldsymbol{r}$ is identical, and approaches 1 as the returns become more dissimilar. In the special case where the covariance matrix is the identity, the function of cosine similarity reduces to

$$\sqrt{1-S_c(\boldsymbol{r},\boldsymbol{e})^2} = \sqrt{1-\frac{\left(\sum_{i=1}^{N} r_i\right)^2}{N\sum_{i=1}^{N} r_i^2}} \in [0,1)$$

This Euclidean special case of the spread measure has certain key differences from a more common measure of spread, the standard deviation. First, it is in the range $[0,1)$, whereas standard deviation is unbounded in the range $[0,\infty)$. Second, this measure is scale-free, whereas any linear transformation of $\boldsymbol{r}$ would scale the standard deviation. Third, it does not perform centering as does standard deviation, which centers by the mean. Because of these characteristics, the function of cosine similarity is sensitive to outliers, and effectively measures spread by distances between data clusters. A higher value for this function of cosine similarity would imply that there is a collection of assets in the market that have significantly different returns from the others.

We can therefore see that $u$ measures the value of using the set of assets returns in mean-variance optimization. In a market with a high Mahalanobis distance, where the asset returns are statistically significantly distant from zero, investors will achieve greater return per unit of risk accounting for standard deviations and cross-correlations in a diversified portfolio. As a simple illustrative example, a Mahalanobis distance with two assets with similar expected returns and high correlation will be less than a Mahalanobis distance between two assets with similar expected returns and low correlation. The two assets with low correlation have greater use in a diversified portfolio as measured by the Mahalanobis distance. When the cosine similarity spread is 0, which implies that the assets all have the same expected return, the efficient frontier would only consist of the minimum variance portfolio, as any allocation of weights will yield the same expected portfolio return. When all the assets have the same expected return except for one, the cosine similarity spread is dependent on how different the expected returns of the asset with different expected return and the other assets. In this case, the portfolio will reduce to a two-asset portfolio, as the only two assets that will be selected are the asset with different expected return and the asset with least covariance to that asset. When the cosine similarity spread is greater, the set of assets will not reduce, and more of the assets will be useful for mean-variance optimization.

## FORECASTING DIRECTIONAL MARKET RETURN

### Online CART

To develop the forecasting model, we first feature engineered the one-month lagged efficient frontier coefficients for each month as our predictor data, and split the data into a train and test set. To forecast each out-of-sample month, we trained the CART model on the current in-sample data.

The main CART hyperparameter that controls overfitting is the max depth. As the max depth grows, the tree is more prone to overfit. Since there are only 3 features, we limited the depth to either 1 or 2, and selected among these two options with a cross-validated grid search. To select the max depth hyperparameter, we elected to use a time-series split rather than standard k-fold cross validation because of the temporal nature of the data. A time-series split ensures that there is no look-ahead bias by partitioning the train sets to include the first k folds in the kth split, and the validation sets to be the (k+1)th fold. As there are only about 300 total monthly data points, we elected to use only 5 folds. We selected accuracy as the metric to maximize in cross-validation.

To account for the loss of information from binarizing the monthly returns, we provided sample weights to the CART model with weights equal to the absolute value of the return, representing how informative that data point is. After forecasting each month's directional market return, we walk the CART model forward by observing the current month's market return, and then repeat the training and forecasting process. Exhibit 1 shows the of the decision tree that was generated for the first out-of-sample period using this methodology.

[EXHIBIT 1 HERE]

An advantage of using CART to forecast the market return is in its interpretability. The selected decision tree can used for post-hoc analysis by providing a visual of how the model uses the features to forecast the market direction, which can be used to explain why the model made certain forecasts.

### Preventing Backtest Overfit

Aldridge et al. (2019) warn of the dangers of backtest overfit using an example of a spurious strategy that trades stocks based on the first letter of their names, and yields high performance for a long backtest period. They explain that modelling financial data with machine learning is especially prone to the curse of dimensionality, as the number of features is often particularly large relative to the number of samples. Researchers often test multiple hypotheses and only keep the best performing results, which are often spurious.

To combat these issues, we limit our feature set to only the three efficient frontier coefficients, as compared to other researchers that use over a hundred features. We also only test the hypothesis that the direction of the market is predictable with a nonlinear relationship to the efficient frontier coefficients. Rather than trying many different models, we limit the potential models to only an online CART, as it is interpretable and can capture nonlinear trends. As CART models are prone to overfit, we limited the max depth to either 1 or 2, as opposed to other researchers that allow decision tree models to go beyond depth 10. We limited the variations of our model to trying only with and without sample weights, and with a time-series split instead of a k-fold cross-validation.

### Forecast Performance Evaluation

We evaluated several performance metrics of the online CART forecasts against multiple baselines as shown in Exhibit 2. As a baseline, we compared this to the online

CART trained on two other feature sets: 5 common technical indicators (Stochastic Oscillator K, Momentum, RSI, A/D Oscillator, and CCI), and the Fama-French 3-factors (market minus risk-free, small minus big, and high minus low) (Fama and French 1993).

[EXHIBIT 2 HERE]

The CART forecast using efficient frontier coefficients achieve better accuracy, precision, recall, F1 (the harmonic mean of precision and recall), and Negative Predictive Value (NPV) than the baselines. The CART forecast investment strategy using the Fama-French 3-factors did not achieve high performance because the factors were proposed to be used for regression, and not forecasting. Most notably among the metrics, the two baselines have NPV below 0.5, so they perform worse than chance at predicting the market being down, while the efficient frontier coefficients has a 65% NPV.

## INTEGRATING FORECASTS TO PORTFOLIO OPTIMIZATION

To explain the process of integrating these forecasts to a diversified portfolio investment strategy, we will use the following notation:

$\tilde{t}$ = the next time period from the current training period

$\boldsymbol{X}_{\tilde{t}-1}(t),\ t \in [0,\ \dots,\ \tilde{t}-1]$ = train efficient frontier coefficient data

$\widetilde{\boldsymbol{X}}_{\tilde{t}-1}(t)\ ,\ t \in [\tilde{t},\ \dots,\ T]$ = test efficient frontier coefficient data

$y_{\tilde{t}-1}(t),\ t \in [0,\ \dots,\ \tilde{t}-1]$ = train market return data

$\widetilde{\boldsymbol{X}}_{\tilde{t}-1}(t)\ ,\ t \in [\tilde{t},\ \dots,\ T]$ = test efficient frontier coefficient data

$r_a$ = the log return of an asset

$r_m$ = the log return of the market

We start with our directional forecasts

$$CART_{\boldsymbol{X}_{\tilde{t}-1}, y_{\tilde{t}-1}}\left(\tilde{X}_{\tilde{t}-1}(\tilde{t}), \theta\right) = \hat{p}_{\mathrm{r_m}>0,t}$$

Under CAPM, we have

$$\mathbb{E}[r_a] = \beta_a\left(\mathbb{E}[r_m] - r_f\right) + r_f$$

$$R_a^2 = \frac{SSR_{a,m}}{SST_{a,m}}$$

Using CAPM, we can find the expected return of an asset conditional on the market forecast.

$$\mathbb{E}\left[r_a|\hat{p}_{\mathrm{r_m}>0}\right] = \mathbb{E}\left[r_a|\ \Pr(r_m > 0) = \hat{p}_{\mathrm{r_m}>0}\right] = \beta_a\left(\mathbb{E}\left[r_m|\ \Pr(r_m > 0) = \hat{p}_{\mathrm{r_m}>0}\right] - r_f\right) + r_f$$

To determine the conditional expectations of the market, we can use a generalization of the inverse Mills ratio. The conditional expected return is the expected return modified by the standard deviation scaled by the ratio of the standard normal density and standard normal cumulative distribution function, which is known as the inverse Mills ratio (Greene 2003). This property of the Mills ratio for a truncated normal distribution is often utilized in regression to account for selection bias from a censored variable (Greene 2003). Assuming the log returns, $r_m \sim N(\mu_m, \sigma_m^2)$, then with the inverse Mills ratio property, we have

$$\mathbb{E}[r_m | r_m > 0] = \mu_m + \sigma_m \frac{\phi\left(\frac{-\mu_m}{\sigma_m}\right)}{1 - \Phi\left(\frac{-\mu_m}{\sigma_m}\right)}$$
$$\mathbb{E}[r_m | r_m < 0] = \mu_m - \sigma_m \frac{\phi\left(\frac{-\mu_m}{\sigma_m}\right)}{\Phi\left(\frac{-\mu_m}{\sigma_m}\right)} \quad (5)$$

We can generalize this conditional expectation with the probability of the market return being greater than 0.

$$\mathbb{E}[r_m | \Pr(r_m > 0)] = \mathbb{E}\left[r_m \middle| \Pr(r_m > 0) = \hat{p}_{\mathrm{r_m>0}}\right]$$

See Appendix B for the derivation.

$$\mathbb{E}\left[r_m \middle| \hat{p}_{\mathrm{r_m>0}}\right] = \mu_m + \sigma_m \frac{(2\hat{p}_{\mathrm{r_m>0}} - 1)\phi\left(\frac{-\mu_m}{\sigma_m}\right)}{\hat{p}_{\mathrm{r_m>0}} - (2\hat{p}_{\mathrm{r_m>0}} - 1)\Phi\left(\frac{-\mu_m}{\sigma_m}\right)} \quad (6)$$

When $\hat{p}_{\mathrm{r_m>0}}$ is 0 or 1, the conditional expectation degenerates to Equation (5) With the conditional expectation of the market, we now have

$$\mathbb{E}\left[r_a | \hat{p}_{\mathrm{r_m>0}}\right] = \beta_a \left( \mu_m + \sigma_m \frac{(2\hat{p}_{\mathrm{r_m>0}} - 1)\phi\left(\frac{-\mu_m}{\sigma_m}\right)}{\hat{p}_{\mathrm{r_m>0}} - (2\hat{p}_{\mathrm{r_m>0}} - 1)\Phi\left(\frac{-\mu_m}{\sigma_m}\right)} - r_f \right) + r_f \quad (7)$$

Since CAPM often does not hold in practice, for our estimation of the return vector, we weight the conditional expectations by the $R^2$ of each CAPM regression.

$$\bar{r}_{\tilde{t}}^{*} = R_a^2 \mathbb{E}[r_a | \hat{y}_{\tilde{t}}] + (1 - R_a^2)\mathbb{E}[r_a], \forall a \in A \quad (8)$$

Because the probabilistic forecasts are often near 0.5, which provides little information, we raised the confidence on our model by discretizing the probabilistic forecasts to either 0 or 1. We now use this as our estimate of the return vector, and calculate the tangency portfolio.

To summarize the entire process, Exhibit 3 provides a visualization of the model process, with our novel contributions in orange.

[EXHIBIT 3 HERE]

## EMPIRICAL ANALYSIS

### Ensuring Realistic Portfolios

To construct more realistic simulated portfolios, we limited leverage and added 1% transaction fees. To ensure that the tangency portfolios did not transfer all of the risk toward leverage, we limited the gross market exposure to 1.5 so that the portfolios are no more than 1.5× levered. We enforce this by adding a constraint to the mean-variance optimization.

We also added 1% daily transaction fees by deducting 1% of the sum of the absolute daily differences in asset weights. We selected 1% for the transaction fees to reflect the commissions of large brokerages, and given that this model would be used mainly on large and liquid assets, the bid/ask spread will be narrow, and will be covered by the 1% transaction fee.

**Empirical Results**

The CART was initially trained on data from 1999-2007, and the portfolio returns were measured out-of-sample from 2008-2022. The online CART forecasted each month, and the portfolios were rebalanced monthly. We measured the performance of our proposed model against three benchmarks: The tangency portfolio rebalanced monthly, the equal-weighed portfolio, and the S&P 500. Exhibit 4 shows that the proposed model outperforms all the benchmarks across the 15 years.

[EXHIBIT 4 HERE]

The proposed model has a higher Sharpe ratio as compared to the three of the benchmark portfolios as shown in Exhibit 5, with high annual return relative to the max drawdown. The proposed model also has statistically significant alpha relative to the three benchmarks as shown in Exhibit 6.

[EXHIBIT 5 HERE]

[EXHIBIT 6 HERE]

These results have theoretical implication for the market, suggesting that the efficient frontier coefficients can define a tree structure that describes the directional return of the market.

## CONCLUSIONS AND FUTURE RESEARCH

This paper presents a novel asset allocation approach using a market forecast with a CART model trained on efficient frontier coefficients. This method has implications on the structure of the directional returns of the market, suggesting that it follows a tree structure defined by efficient frontier coefficients. This method is interpretable, as it allows users to observes the underlying tree model, and does not require additional input estimates like the Black-Litterman model.

Our proposed method makes use of the decomposition of efficient frontiers into three interpretable coefficients: $r_{MVP}$, $\sigma_{MVP}$, and $u$. The proposed method uses an online CART model trained on the feature-engineered interpretable set of efficient frontier coefficients. The method then uses the expectations of the asset returns conditional on the market forecast as the parameter estimate for the return vector, then invests at the tangency portfolio with this return estimate. The conditional expected return is calculated using CAPM and the inverse Mills ratio.

This method is more robust than the standard mean-variance model as the market forecast implied return vector provides a better estimate. Using this method out-of-sample from 2000-2022, this model outperformed three benchmark portfolios: the tangency portfolio, the equal-weighted, and the S&P 500. The proposed model achieved a higher Sharpe ratio as compared to these benchmarks.

Although this work yielded significant results, there still are limitations that we will investigate in future research. We will test the model on other sets of assets that span the market than the sector ETFs, as these results could be dependent on the selected assets. While the CART model trained on the efficient frontier coefficients data outperformed the model trained on technical indicators, there may be a superior model with some combination of these sets of features. We will investigate combining these feature sets and trying more sophisticated classification models such as random forests and SVMs.

**REFERENCES**

Aldridge, Irene, and Marco Avellaneda. 2019. "Neural Networks in Finance: Design and Performance." *The Journal of Financial Data Science* 1, no. 4 (Fall): 39-62.

Alexander, Nolan, William Scherer, and Matt Burkett. 2021. "Extending the Markowitz Model with Dimensionality Reduction: Forecasting Efficient Frontiers." *Proceedings of the 2021 IEEE Systems and Information Engineering Design Symposium (SIEDS), Charlottesville, VA,* 29-30.

Alexander, Nolan, and William Scherer. 2023. "Forecasting Tangency Portfolios and Investing in the Minimum Euclidean Distance Portfolio to Maximize Out-of-Sample Sharpe Ratios." *MDPI Engineering Proceedings of the 2023 International conference on Time Series and Forecasting (ITISE), Gran Canaria, Spain*.

Arnott, Rob, Harvey, Campbell, and Harry Markowitz. 2019. "A Backtesting Protocol in the Era of Machine Learning." *The Journal of Financial Data Science* 1, no. 1 (Winter): 64-74.

Black, Fischer and Robert Litterman. 1992. "Global Portfolio Optimization." *Financial Analysts Journal* 48 (5): 28–43.

Cont, Rama,. 2001. "Empirical Properties of Asset Returns: Stylized Facts and Statistical Issues." *Quantitative Finance* 1 (2): 223–36.

Fama, Eugene F., and Kenneth R. French. 1993. "Common Risk Factors in the Returns on Stocks and Bonds." *Journal of Financial Economics* 33, no. 1 (February): 3–56.

Greene, William H. 2003. *Econometric Analysis*. New Jersey: Prentice-Hall.

French, Kenneth R. "Kenneth R. French Data Library." Accessed 3 March 2021. https://mba.tuck.dartmouth.edu/pages/faculty/ken.french/data_library.html.

Kim, Hongjoong, Sookyung Jun and Kyoung-Sook Moon. 2022. "Stock Market Prediction Based on Adaptive Training Algorithm in Machine Learning." *Quantitative Finance* 22, no. 6 (February): 1–20.

Krauss, Christopher, Xuan A. Do and Nicolas Huck. 2017. "Deep Neural Networks, Gradient-Boosted Trees, Random Forests: Statistical Arbitrage on the S&P 500." *European Journal of Operational Research* 259, no. 2 (June): 689–702.

Kumar, Manish, and M Thenmozhi. 2005. "Forecasting Stock Index Movement: A Comparison of Support Vector Machines and Random Forest." *Proceedings of 9th Indian Institute of Capital Markets Conference*.

Li, Yimou, Zachary Simon, and David Turkington. 2021. "Investable and Interpretable Machine Learning for Equities." *The Journal of Financial Data Science* 4, no. 1 (Winter): 54-74.

Luenberger, David. 1997. *Investment Science*. New York, NY: Oxford University Press.

Markowitz, Harry. 1952. "Portfolio Selection." *The Journal of Finance* 7, no. 1 (March): 77.

Merton, Robert C. 1972. "An Analytic Derivation of the Efficient Portfolio Frontier." *The Journal of Financial and Quantitative Analysis* 7, no. 4 (September): 1851.

Patel, Jigar, Sahil Shah, Priyank Thakkar, and Ketan Kotecha. 2015. "Predicting Stock and Stock Price Index Movement Using Trend Deterministic Data Preparation and Machine Learning Techniques." *Expert Systems with Applications* 42, no. 1 (January): 259–68.

Petters, Arlie O., and Xiaoying Dong. 2016. *An Introduction to Mathematical Finance with Applications: Understanding and Building Financial Intuition*, Switzerland: Springer.

Rapach, David, Jack Strauss, Jun Tu, and Guofu Zhou. 2019. "Industry Return Predictability: A Machine Learning Approach." *The Journal of Financial Data Science* 1, no. 3 (Summer): 9-28.

Sharpe, Willaim F. 1964. "Capital Asset Prices: A Theory of Market Equilibrium Under Conditions of Risk." *The Journal of Finance* 19, no. 3 (September): 425–42.

Sharpe, William F. 1966. "Mutual Fund Performance." *The Journal of Business*, 39, no. 1 (January): 119–38.

Sirignano, Justin, and Rama Cont 2019. "Universal Features of Price Formation in Financial Markets: Perspectives from Deep Learning." *Quantitative Finance* 19, no. 9 (November): 1449–59.

Tobin, James. 1958. "Liquidity Preference as Behavior Towards Risk." *The Review of Economic Studies*, 25, no. 2 (February): 65.

Wang, Haifeng, Harshdeep Ahluwalia., Roger Aliaga-Díaz, and Joseph Davis. 2021. "The Best of Both Worlds: Forecasting US Equity Market Returns Using a Hybrid Machine Learning–Time Series Approach." *The Journal of Financial Data Science* 3, no. 2 (Spring): 9-20.

## EXHIBITS

Exhibit 1. The decision tree generated with the efficient frontier coefficients

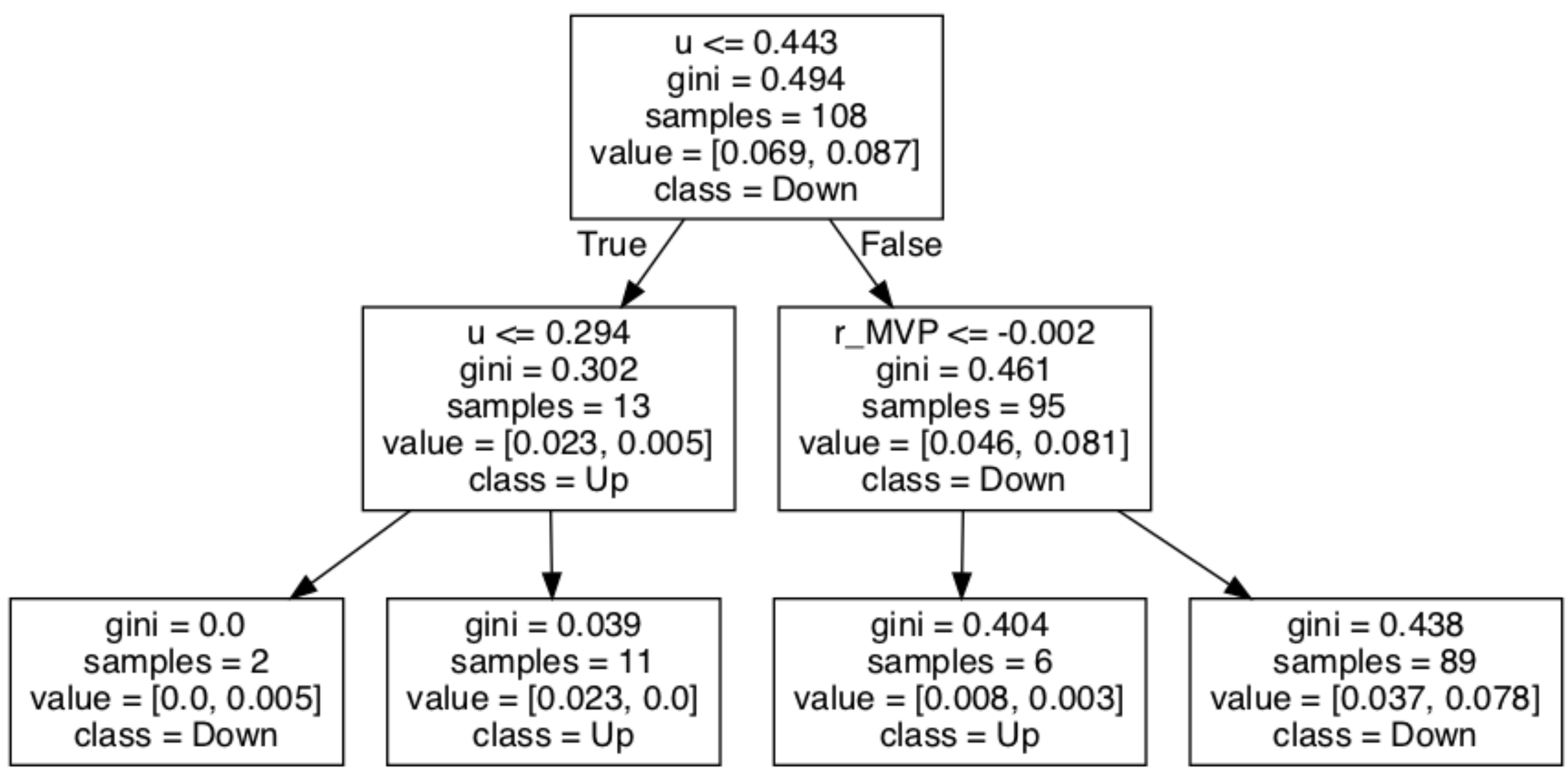

**Exhibit 2. Online CART Forecast Metrics**

| **Features** | **Accuracy** | **Precision** | **Recall** | **F1** | **NPV** |
|---|---|---|---|---|---|
| **Efficient Frontier Coefs** | 69% | 69% | 95% | 80% | 65% |
| **Tech Indicators** | 63% | 66% | 91% | 76% | 35% |
| **Fama-French Factors** | 61% | 66% | 87% | 75% | 32% |

Exhibit 3. A visualization of the entire process of the proposed model

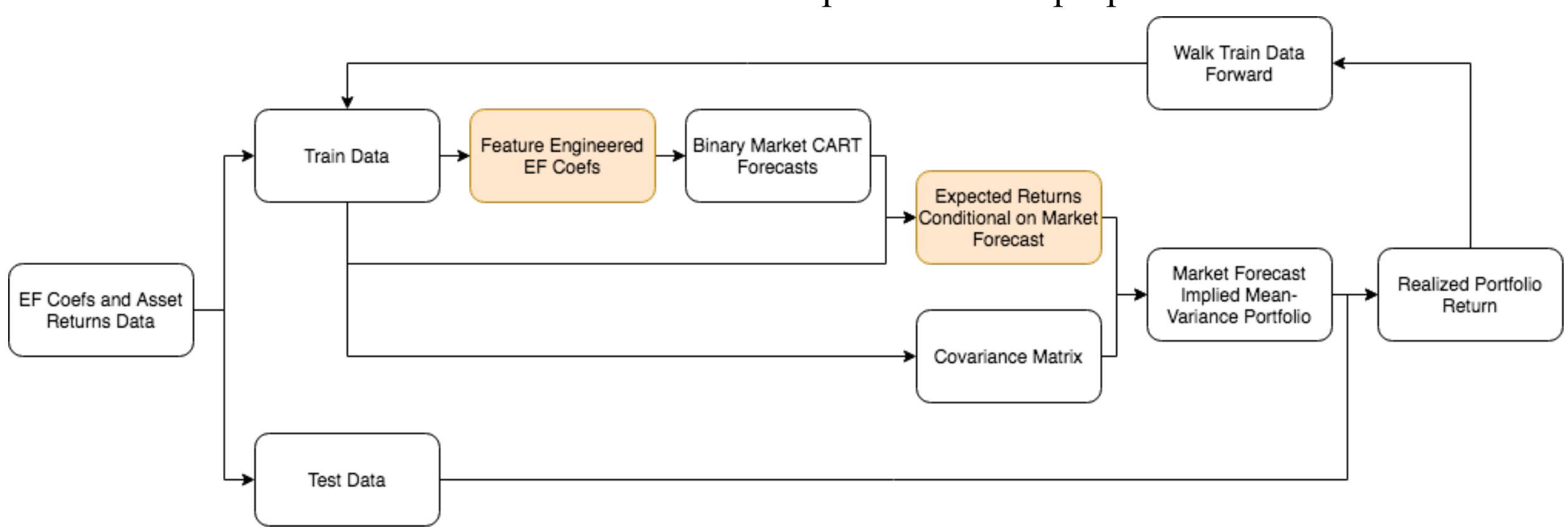

Exhibit 4. The performance of the CART market forecast portfolio as compared to three benchmarks

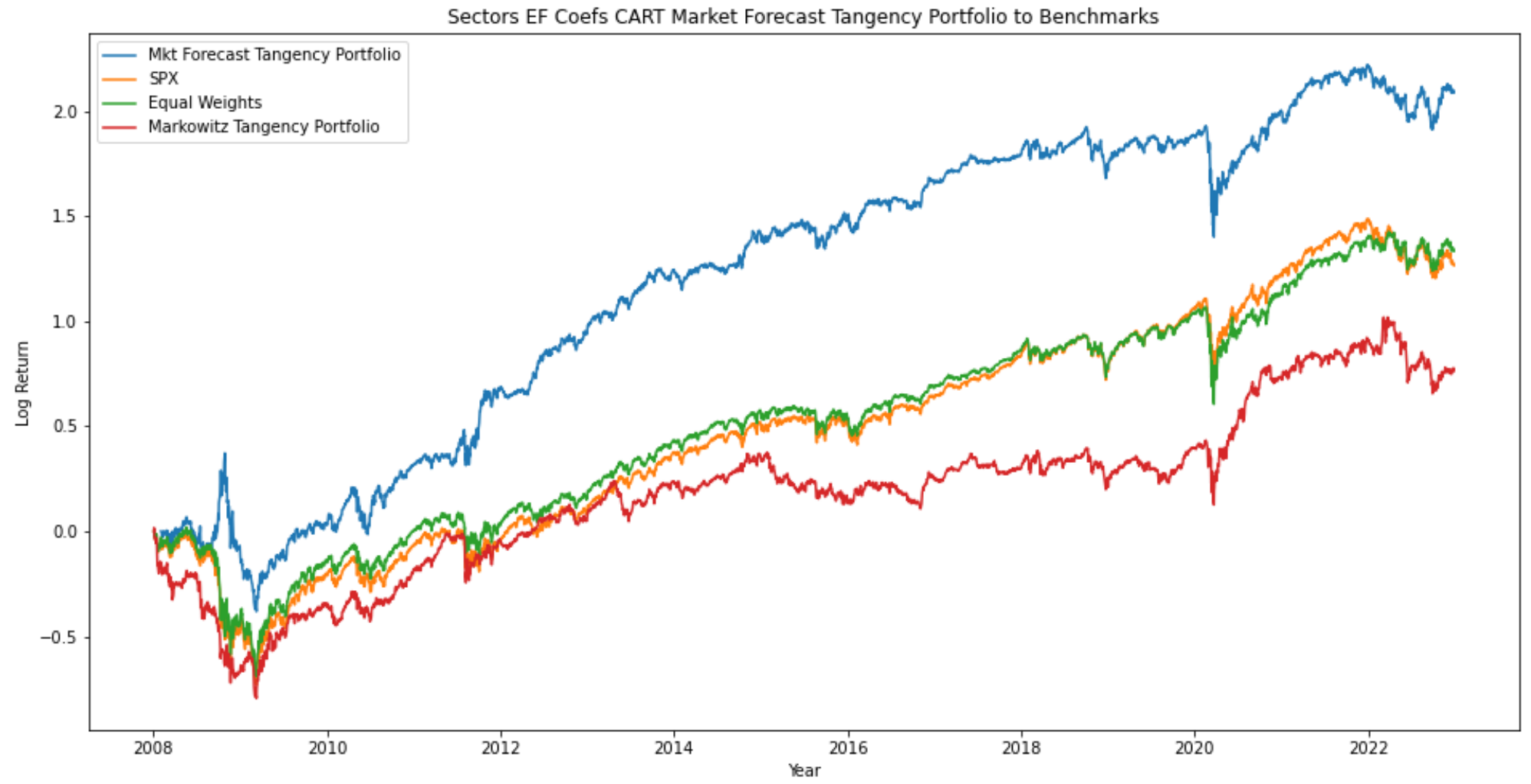

| Exhibit 5. Portfolio Metrics from 2008 to 2022 | | | |
|---|---|---|---|
| **Portfolio** | **Sharpe Ratio** | **Annual Return** | **Max Drawdown** |
| **CART Tangency Weights Portfolio** | 0.73 | 17.8% | -53% |
| **Monthly Tangency Portfolio** | 0.33 | 7.4% | -56% |
| **Equal Weighted** | 0.52 | 11.6% | -51% |
| **S&P 500** | 0.49 | 11.2% | -52% |

| Exhibit 6. CART Portfolio Alpha Regressions | | |
|---|---|---|
| **Portfolio** | **Alpha** | **P-value** |
| **Monthly Tangency Portfolio** | 11.6% | 0.008 |
| **Equal Weighted** | 10.5% | 0.02 |
| **S&P 500** | 9.9% | 0.02 |

## APPENDIX A: Derivation of Alternate Form of $u$

We first decompose $u$ into Cauchy-Schwarz form using Eq. (4) and Eq. (1). Define the $\boldsymbol{V}^{-1}$-inner product as $\langle \boldsymbol{x}, \boldsymbol{y} \rangle_{\boldsymbol{V}^{-1}} = \boldsymbol{x}^T \boldsymbol{V}^{-1} \boldsymbol{y}$, with induced norm $\| \boldsymbol{x} \|_{\boldsymbol{V}^{-1}} = \sqrt{\boldsymbol{x}^T \boldsymbol{V}^{-1} \boldsymbol{x}}$. This is a valid inner product because $\boldsymbol{V}^{-1}$ is symmetric and positive definite. With this notation, the Merton coefficients become $A = \| \boldsymbol{e} \|_{\boldsymbol{V}^{-1}}^2$, $B = \langle \boldsymbol{r}, \boldsymbol{e} \rangle_{\boldsymbol{V}^{-1}}$, and $C = \| \boldsymbol{r} \|_{\boldsymbol{V}^{-1}}^2$.

Starting from Eq. (4):

$$
\begin{aligned}
u^2 &= \frac{AC - B^2}{A} \\
&= \frac{\| \boldsymbol{e} \|_{\boldsymbol{V}^{-1}}^2 \| \boldsymbol{r} \|_{\boldsymbol{V}^{-1}}^2 - \langle \boldsymbol{r}, \boldsymbol{e} \rangle_{\boldsymbol{V}^{-1}}^2}{\| \boldsymbol{e} \|_{\boldsymbol{V}^{-1}}^2}
\end{aligned}
$$

We can now see that $u \geq 0$ because $\| \boldsymbol{e} \|_{\boldsymbol{V}^{-1}}^2 \| \boldsymbol{r} \|_{\boldsymbol{V}^{-1}}^2 - \langle \boldsymbol{r}, \boldsymbol{e} \rangle_{\boldsymbol{V}^{-1}}^2 \geq 0$ by the Cauchy-Schwarz inequality applied to $\langle \cdot, \cdot \rangle_{\boldsymbol{V}^{-1}}$, and $\| \boldsymbol{e} \|_{\boldsymbol{V}^{-1}}^2 > 0$ because $\boldsymbol{V}^{-1}$ is positive definite.

Factoring out $\| \boldsymbol{e} \|_{\boldsymbol{V}^{-1}}^2$ from the numerator:

$$u^2 = \| \boldsymbol{r} \|_{\boldsymbol{V}^{-1}}^2 - \frac{\langle \boldsymbol{r}, \boldsymbol{e} \rangle_{\boldsymbol{V}^{-1}}^2}{\| \boldsymbol{e} \|_{\boldsymbol{V}^{-1}}^2}$$

$$= \| \boldsymbol{r} \|_{\boldsymbol{V}^{-1}}^2 \left( 1 - \frac{\langle \boldsymbol{r}, \boldsymbol{e} \rangle_{\boldsymbol{V}^{-1}}^2}{\| \boldsymbol{e} \|_{\boldsymbol{V}^{-1}}^2 \| \boldsymbol{r} \|_{\boldsymbol{V}^{-1}}^2} \right)$$

The cosine similarity in the $\boldsymbol{V}^{-1}$-inner product space is defined as

$$S_c^{(\boldsymbol{V}^{-1})}(\boldsymbol{r}, \boldsymbol{e}) = \frac{\langle \boldsymbol{r}, \boldsymbol{e} \rangle_{\boldsymbol{V}^{-1}}}{\| \boldsymbol{r} \|_{\boldsymbol{V}^{-1}} \| \boldsymbol{e} \|_{\boldsymbol{V}^{-1}}}$$

$$= \frac{\boldsymbol{r}^T \boldsymbol{V}^{-1} \boldsymbol{e}}{\sqrt{(\boldsymbol{e}^T \boldsymbol{V}^{-1} \boldsymbol{e})(\boldsymbol{r}^T \boldsymbol{V}^{-1} \boldsymbol{r})}}$$

Substituting, we obtain

$$u^2 = \| \boldsymbol{r} \|_{\boldsymbol{V}^{-1}}^2 \left( 1 - S_c^{(\boldsymbol{V}^{-1})}(\boldsymbol{r}, \boldsymbol{e})^2 \right)$$

Therefore

$$u = \| \boldsymbol{r} \|_{\boldsymbol{V}^{-1}} \cdot \sqrt{1 - S_c^{(\boldsymbol{V}^{-1})}(\boldsymbol{r}, \boldsymbol{e})^2}$$

$$= \sqrt{\boldsymbol{r}^T \boldsymbol{V}^{-1} \boldsymbol{r}} \cdot \sqrt{1 - S_c^{(\boldsymbol{V}^{-1})}(\boldsymbol{r}, \boldsymbol{e})^2}$$

Therefore, $u$ can be defined by the product of the Mahalanobis distance of $\boldsymbol{r}$ to $\boldsymbol{0}$, which is $\| \boldsymbol{r} \|_{\boldsymbol{V}^{-1}} = \sqrt{\boldsymbol{r}^T \boldsymbol{V}^{-1} \boldsymbol{r}}$, and a function of the cosine similarity in the $\boldsymbol{V}^{-1}$-weighted inner product space, $\sqrt{1 - S_c^{(\boldsymbol{V}^{-1})}(\boldsymbol{r}, \boldsymbol{e})^2}$. Note that this cosine similarity is computed in the $\boldsymbol{V}^{-1}$-inner product space rather than the standard Euclidean space, so it accounts for the covariance structure of the assets. When $\boldsymbol{V} = \boldsymbol{I}$, this reduces to the standard Euclidean cosine similarity.

## APPENDIX B: Derivation of the Conditional Expectation Given a Probabilistic Forecast of Direction

$$\mathbb{E}[r_m | \Pr(r_m > 0)] = \mathbb{E}\left[ r_m \middle| \Pr(r_m > 0) = \hat{p}_{\mathrm{r_m}>0} \right]$$

$$= \mathbb{E}\left[ \mu_m + \sigma_m Z \middle| \Pr\left( Z > \frac{0 - \mu_m}{\sigma_m} \right) = \hat{p}_{\mathrm{r_m}>0} \right]$$

$$= \mu_m + \sigma_m \mathbb{E}\left[ Z \middle| \Pr\left( Z > \frac{-\mu_m}{\sigma_m} \right) = \hat{p}_{\mathrm{r_m}>0} \right]$$

$$= \mu_m + \sigma_m \frac{\hat{p}_{\mathrm{r_m}>0} \int_{\frac{-\mu_m}{\sigma_m}}^{\infty} z\phi(z) + \left(1 - \hat{p}_{\mathrm{r_m}>0}\right) \int_{-\infty}^{\frac{-\mu_m}{\sigma_m}} z\phi(z)}{\hat{p}_{\mathrm{r_m}>0} \Pr\left( Z > \frac{-\mu_m}{\sigma_m} \right) + \left(1 - \hat{p}_{\mathrm{r_m}>0}\right) \Pr\left( Z < \frac{-\mu_m}{\sigma_m} \right)}$$

$$= \mu_m + \sigma_m \frac{\hat{p}_{\mathrm{r_m}>0} \int_{\frac{-\mu_m}{\sigma_m}}^{\infty} -\phi'(z)dz + \left(1 - \hat{p}_{\mathrm{r_m}>0}\right) \int_{-\infty}^{\frac{-\mu_m}{\sigma_m}} -\phi'(z)dz}{\hat{p}_{\mathrm{r_m}>0} \left( 1 - \Phi\left( \frac{-\mu_m}{\sigma_m} \right) \right) + \left(1 - \hat{p}_{\mathrm{r_m}>0}\right) \Phi\left( \frac{-\mu_m}{\sigma_m} \right)}$$

$$= \mu_m + \sigma_m \frac{\hat{p}_{\mathrm{r_m}>0} \phi\left( \frac{-\mu_m}{\sigma_m} \right) + \left(1 - \hat{p}_{\mathrm{r_m}>0}\right) \left( -\phi\left( \frac{-\mu_m}{\sigma_m} \right) \right)}{\hat{p}_{\mathrm{r_m}>0} - \left(2\hat{p}_{\mathrm{r_m}>0} - 1\right) \Phi\left( \frac{-\mu_m}{\sigma_m} \right)}$$

$$\mathbb{E}\left[r_m \middle| \hat{p}_{\mathrm{r_m>0}}\right] = \mu_m + \sigma_m \frac{(2\hat{p}_{\mathrm{r_m>0}} - 1)\phi\left(\frac{-\mu_m}{\sigma_m}\right)}{\hat{p}_{\mathrm{r_m>0}} - (2\hat{p}_{\mathrm{r_m>0}} - 1)\Phi\left(\frac{-\mu_m}{\sigma_m}\right)} \tag{8}$$